\documentclass[fleqn,twoside]{article}
\usepackage{espcrc2}

\usepackage{graphicx}
\usepackage{epsfig}
\usepackage[figuresright]{rotating}

\newcommand{\AmS}{{\protect\the\textfont2
  A\kern-.1667em\lower.5ex\hbox{M}\kern-.125emS}}

\title{Multi-lepton Production at High Transverse Momentum \\ in $ep$ collisions at HERA}

\author{Cristinel Diaconu\address[CPPM]{Centre de Physique des Particules de Marseille,\\
        163 Avenue de Luminy, 
        13288 Marseille cedex 09, France} \\on behalf of H1 and ZEUS Collaborations%
}
       
\begin{document}

\begin{abstract}
Multi-electron and multi-muon production have been measured at high transverse momentum
in  electron-proton collisions   at HERA.
Good overall agreement is found with the Standard Model predictions,
dominated by photon-photon interactions. Events are observed with
 a di-electron mass above 100 GeV, a domain where the Standard Model
 prediction is low.
\vspace{1pc}
\end{abstract}

\maketitle

\section{Introduction}

\renewcommand{\tabcolsep}{1.1pc} %
\renewcommand{\arraystretch}{1.05} %

\begin{table*}[htb]
\begin{center}
 
\begin{tabular}{|l|c|c||c|c|}\hline
   Selection   & DATA  &  SM &  GRAPE  &  NC-DIS + Compton \\ \hline \hline
\multicolumn{1}{|l}{\bf H1 115 pb$^{\mathrm \bf -1}$}& \multicolumn{4}{l|}{$20^\circ < \theta^{e1,2} < 150^\circ$, $P_T^{e1}>10$~GeV, $P_T^{e2}>5$~GeV }\\ 
\hline 
  Visible 2e           &   105 &  $118.2\pm  12.8$ &  $ 93.3\pm  11.5$ &  $ 25.0\pm   5.5$ \\
  Visible 3e           &    16 &  $ 21.6\pm   3.0$ &  $ 21.5\pm   3.0$ &  $  0.1\pm   0.1$ \\
\hline \hline
 
 Visible 2e $M_{12}>100$~GeV &   3 &  $ 0.25\pm  0.05$ &  $ 0.21\pm  0.04$ &  $ 0.04\pm  0.03$ \\
 Visible 3e $M_{12}>100$~GeV &   3 &  $ 0.23\pm  0.04$ &  $ 0.23\pm  0.04$ &  $ 0.00\pm  0.00$ \\
   \hline
\hline 
\multicolumn{1}{|l}{\bf ZEUS 130 pb$^{\mathrm \bf -1}$}&\multicolumn{4}{l|}{$17^\circ < \theta^{e1,2} < 167^\circ$, $P_T^{e1}>10$~GeV, $E^{e2}>10$~GeV }\\ 
\hline
  Visible 2e           &   191 &  $213.9\pm  3.9$ &  $ 182.2\pm  1.2$ &  $ 31.7\pm   3.7$ \\
  Visible 3e           &    26 &  $ 34.7\pm  0.5$ &  $ 34.7\pm   0.5$ &  $-$ \\
\hline \hline
 
Visible 2e $M_{12}>100$~GeV &   2 &  $ 0.77\pm  0.08$ &  $ 0.47\pm  0.05$ &  $ 0.30\pm  0.07$ \\
 Visible 3e $M_{12}>100$~GeV &   0 &  $ 0.37\pm  0.04$ &  $ 0.37\pm  0.04$ &  $-$ \\
   \hline
 \end{tabular}
 
\caption{ Observed and predicted multi-electron event rates for all selected events and for events with masses $M_{12}>100$~GeV as function of the number of identified electrons. The prediction errors for the H1 analysis include model uncertainties and experimental systematical errors added in quadrature. For ZEUS analysis, the predicted rates are shown with the statistical errors of the Monte Carlo only.}

\label{tab:h1zeusme}
\end{center}
\end{table*}

\begin{figure*}[htb]
\begin{center}
  \epsfig{file=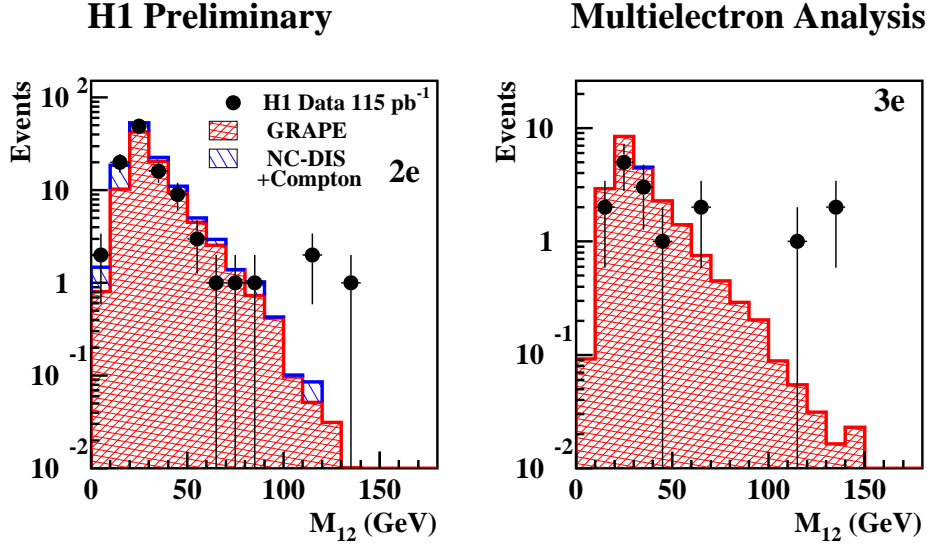,width=140mm}%
\end{center}
  \caption{ Distribution of the invariant mass $M_{12}$ of the two
highest $P_T$ electrons for the H1 analysis.                    
Events classified as di-electrons (left) and tri-electrons (right) are shown.}
\label{fig:h1_mass}
\end{figure*}

The measurement of rare processes provides a unique method to
search for new physics. 
At HERA, two experiments (H1 and ZEUS) study electron-proton collisions with a center of mass energy of 300 to 320 GeV.
In this paper we shortly describe the first measurement
of multi-electron and multi-muon production at high transverse 
momentum ($P_T$)
in $ep$ collisions at HERA~\cite{h1conf,h1confm,zeusconf}. 

\section{Multi-lepton Processes}
Within the Standard Model (SM)
the production of multilepton events in $ep$ collisions is possible
mainly through photon-photon interactions, where quasi-real photons
radiated from the incident electron and proton interact to produce
a pair of leptons $\gamma\gamma\rightarrow \ell^+ \ell^-$~\cite{verm}.
 The GRAPE Monte Carlo generator~\cite{grape} interfaced to
 full detector simulation has been used to simulate multi-lepton production.
 This generator is based on the full electro-weak matrix element calculation
 at tree level. The scattered proton or its remnants is treated 
in three phase space regions: elastic, quasielastic and inelastic. 
\par
The main experimental backgrounds to multi-electron production are
processes where, in addition to a true electron, one or more fake
electrons may be reconstructed from the final state particles.
The dominant contribution is expected from
Neutral Current Deep Inelastic Scattering (NC-DIS) where a fake
electron from the hadrons or a radiated photon is selected together
with the scattered electron. Elastic Compton scattering can also
contribute if the high $P_T$ photon is misidentified as an electron.
Background for multi-muon events are processes where fake muons are reconstructed from isolated hadrons. This contribution is negligible.
\section{Measurement of multi-electron events}
Both the H1 and ZEUS experiments are complex detectors with very good 
lepton identification capabilities. 
The electron identification is based on calorimetric information
 together with tracking conditions for efficient background rejection. 
\par
Electrons are 
measured by both H1 and ZEUS in a large acceptance range $5^\circ < \theta_e < 175^\circ$,
 where $\theta_e$ is the electron polar angle measured with respect to
 the outgoing proton direction.  The electron energy measured from 
calorimetric 
information has to be above 5 GeV. This energy threshold is increased in 
H1 analysis to 10 GeV for electrons candidates with 
 $\theta_e<20^\circ$. The ZEUS analysis requires the electron energy above 10 
GeV for  $\theta_e<164^\circ$.
 The electron candidates have 
to be isolated from other calorimetric deposits.
  In the central region defined by $20^\circ < \theta_e < 150^\circ$ for 
 the H1 analysis and  $17^\circ < \theta_e < 164^\circ$ for the ZEUS 
 analysis, an isolated charged track measured in the central tracking 
 system has to be associated to the calorimetric deposit. The identified 
 electrons  are indexed in decreasing transverse momentum $P_T$: 
$P_T^{e_i} > P_T^{e_{i+1}}$. 
\par
The selection of multi-electron events is based on the requirement of two central electrons with large energy or transverse momentum. Both H1 and ZEUS analysis require the first central electron to have  transverse momentum above 10~GeV. The second central electron is required to have $P_T^{e2}>5$~GeV ($E^{e2}>10$~GeV) in H1 (ZEUS) analysis. Any other electron identified is also counted and the selected events are classified by the number of identified electrons in the event. 
\par
The results of H1 and ZEUS analyses are presented in the table~\ref{tab:h1zeusme}. The H1 analysis on an event sample corresponding to 115 pb$^{-1}$ measured 121 multi-electron events, while ZEUS, with an integrated luminosity of 130 pb$^{-1}$ detected 217 such events. The di-electron sample is dominated by the signal  with a 15-20\% contribution from the background. In the tri-electron sample, the background contribution is negligible.
Both  H1 and ZEUS observations are in good agreement with the predicted 
rates. The main difference between H1 and ZEUS acceptances for the signal is due to different angular range for the central electrons. 
\par
\begin{figure}[hhh]
  \epsfig{file=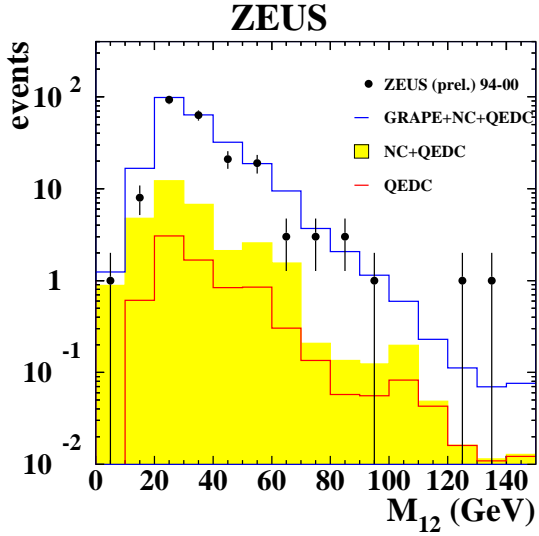,width=75mm}%
\caption{  Distribution of the invariant mass $M_{12}$ of the two
highest $P_T$ electrons for the ZEUS analysis. All observed events with more than two electrons are shown. }
\label{fig:zeus_mass}
\end{figure}

The distributions of the invariant mass of the two highest $P_T$ electrons are shown in figures~\ref{fig:h1_mass}~and~\ref{fig:zeus_mass}. Data are in good overall agreement with the Standard Model prediction. A few events with masses $M_{12}>100$~GeV are observed in a region where the standard model prediction is low. H1 measured 3 di-electron events for 0.25 expected. ZEUS observed 2 di-electron events for 0.77 expected. In the tri-electron sample, H1 observed 3 events with $M_{12}>100$~GeV for an expectation of 0.23 while ZEUS do not observe events in that mass region where the expectation is 0.37. For the high mass di-electron events, the transverse momenta of the two electrons is also large (above 50~GeV). The topology of the observed tri-electron high-mass events is different: the transverse momenta of the two highest $P_T$ electrons is lower (around 30 GeV) and the high mass value is associated with a larger polar opening angle between the two electrons (``forward-backward'' topology). 

\section{Measurement of multi-muon events}
Muons are identified using central tracker reconstructed tracks, calorimetric deposits and muon chamber signals.
A search for multi-muon events has been performed by the H1 and ZEUS experiments in the angular range $20^\circ < \theta < 160^\circ$. At least two muons are required with $P_T^{\mu1,2}>2$~GeV ($P_T^{\mu1,2}>5$~GeV) in H1 (ZEUS) analyses. With an analyzed data sample corresponding to an integrated luminosity of 105 pb$^{-1}$, the ZEUS analysis detects 200 events for an expectation of 213$\pm11_{stat}$. The H1 analysis of an integrated luminosity of 70 pb$^{-1}$ found 1242 multi-muon events in good agreement with the prediction of 1253$\pm125_{stat.\oplus syst.}$. 
\par  
No event with two muons at high mass $M_{\mu\mu}>100$~GeV is observed by either experiment. Starting from the 3 high mass di-electron events observed by H1 the expected observation should be of the order of one di-muon event with $M_{12}>100$~GeV, due to the lower efficiency and luminosity in the multi-muon channel. The expected increase in luminosity at HERA II will improve the knowledge of the high mass region and allow a better comparison between electron and muon channels.
\par
\begin{figure}[hhh]
  \epsfig{file=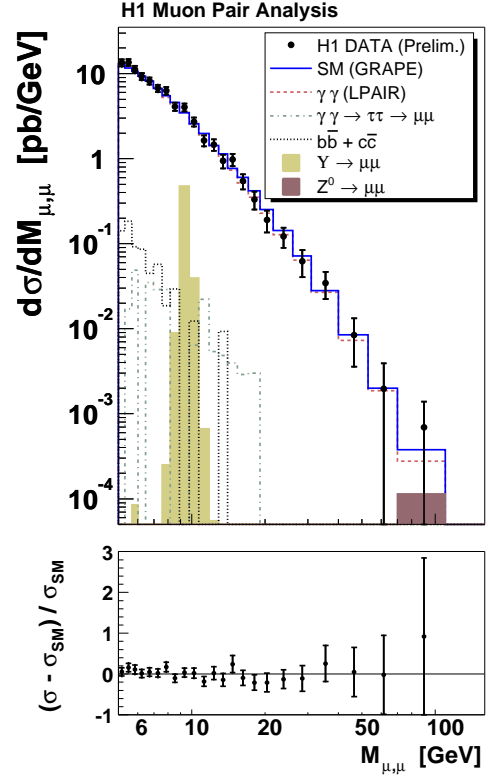,width=75mm}%
\caption{Differential muon pair production cross-section as a function of di-muon mass. The bottom plot shows the relative difference between the data and the Standard Model prediction.}
\label{fig:h1_mumass}
\end{figure}

 Figure~\ref{fig:h1_mumass} presents the visible cross section measured by H1 as a function of the invariant mass of the muon pair compared to the Standard Model prediction. Backgrounds and also other sources of muon pair production like heavy hadron decays are negligible. Very good agreement with the Standard Model prediction is observed up to 80 GeV over a four order of magnitude decrease in the cross section. The integrated cross section in the visible phase space has been measured to be $46.5\pm4.7$~pb in good agreement with the prediction of 46.2~pb. 
\par
The H1 collaboration has measured separate cross-section for inelastic muon pair production. Inelastic events contain hadrons attributed to the proton dissociation  detected in the main calorimeter or in the forward components of the detector. The inelastic cross section has been measured by H1 to be $20.8\pm3.3$~pb in agreement with the expected cross section of 21.5~pb.

\section{Conclusions}

Events with two or three visible leptons(electrons or muons) have been measured for the first time in electron-proton collisions by H1 and ZEUS experiments. Good overall agreement with the Standard Model prediction is observed. In the multi-electron analysis, several events with invariant mass of the two highest $P_T$ electrons $M_{12}>100$~GeV have been observed.  Multi-muon events have also been detected and the differential cross section has been measured and found in very good agreement with the Standard Model. No event with two muons at high mass $M_{\mu\mu}>100$~GeV is observed by either experiment.
The increase in luminosity expected at HERA II  will improve the knowledge of multi-lepton production at high transverse momenta and high mass.

\end{document}